\documentclass[runningheads,a4paper]{llncs}

\usepackage{amssymb}
\usepackage{graphicx}
\usepackage{epsfig}
\usepackage{amsmath}
\usepackage{authblk}
\usepackage{url}
\usepackage{multirow}
\newcommand{\keywords}[1]{\par\addvspace\baselineskip
\noindent\keywordname\enspace\ignorespaces#1}
\usepackage{setspace}

\begin{document}
\begin{spacing}{1.0}

\mainmatter  

\title{BDGS: A Scalable Big Data Generator Suite in Big Data Benchmarking}

\titlerunning{BDGS: A Scalable Big Data Generator Suite in Big Data Benchmarking}

%
%
\author[1,2]{Zijian Ming}
\author[1]{Chunjie Luo}
\author[1,2]{Wanling Gao}
\author[1,3]{Rui Han}
\author[1]{Qiang Yang}
\author[1]{Lei Wang}
\author[1]{Jianfeng Zhan\thanks{The corresponding author is Jianfeng Zhan.}}

\affil[1]{State Key Laboratory Computer Architecture, Institute of Computing Technology, Chinese Academy of Sciences, China}
\affil[2]{University of Chinese Academy of Sciences, China}
\affil[3]{Department of Computing, Imperial College London, UK}

\authorrunning{Zijian Ming and etc.}


\institute{\{mingzijian, luochunjie, gaowanling\}@ict.ac.cn, harryandlina2011@gmail.com, yangqiang@ict.ac.cn,wl@ncic.ac.cn, zhanjianfeng@ict.ac.cn}

%
%

\toctitle{BDGS: A Scalable Big Data Generator Suite in Big Data Benchmark}
\tocauthor{Zijian Ming, Chunjie Luo, Wanling Gao, Rui Han, Qiang Yang, Lei Wang and Jianfeng Zhan}
\maketitle

\vspace{-25pt}

\begin{abstract}
Data generation is a key issue in big data benchmarking that aims to generate application-specific data sets to meet the 4V requirements of big data. Specifically, big data generators need to generate scalable data (Volume) of different types (Variety) under controllable generation rates (Velocity) while keeping the important characteristics of raw data (Veracity). This gives rise to various new challenges about how we design generators efficiently and successfully. To date, most existing techniques can only generate limited types of data and support specific big data systems such as Hadoop. Hence we develop a tool, called Big Data Generator Suite (BDGS), to efficiently generate scalable big data while employing data models derived from real data to preserve data veracity. The effectiveness of BDGS is demonstrated by developing six data generators covering three representative data types (structured, semi-structured and unstructured) and three data sources (text, graph, and table data).

\keywords{Big Data, Benchmark, Data Generator, Scalable, Veracity}
\end{abstract}

\section{Introduction}\vspace{-10pt}
As internet becomes faster, more reliable and more ubiquitous, data explosion is an irresistible trend that data are generated faster than ever. To date, about 2.5 quintillion bytes of information is created per day \cite{dataproduce}. IDC forecasts this speed of data generation will continue and it is expected to increase at an exponential level over the next decade. The above facts mean it becomes difficult to process such huge and complex data sets just using traditional data processing systems such as DBMS. The challenges of capturing, storing, indexing, searching, transferring, analyzing, and displaying \emph{Big Data} bring fast development of big data systems \cite{barroso2009datacenter,zhan2012high,ferdman2011clearing,ghazala,lotfi2012scale}. Within this context, big data benchmarks are developed to address the issues of testing and comparing such systems, thus measuring their performance, energy efficiency, and cost effectiveness.

Big data generation, which aims to generate application-specific data sets for benchmarking big data systems, has become one of the most important features of a benchmark. Generally, using real data sets as workload inputs can guarantee data veracity in benchmarking. Based on our experience, however, we noticed that in many real-world scenarios, obtaining real data sets for benchmarking big data systems is not trivial. First, many owners of real big data sets are not willing to share their data due to confidential issues. Therefore, it is difficult to get real data sets with a variety of types including structured, semi-structured or unstructured data. Moreover, it is difficult to flexibly adapt the volume and velocity of fixed-size real data sets to meet the requirements of different benchmarking scenarios. Finally, the transferring of big data sets via the internet, e.g. downloading TB or PB scale data, is very expensive.

A natural idea to solve these problems is to generate synthetic data used as inputs of workloads on the basis of real-world data. In \cite{gray1994quickly}, Jim Gray discussed how to efficiently generate data using pseudo-random sequence and a non-uniform distribution to test the performance of database systems. In recent years, with the rapid development of big data systems and related technologies, the 4V properties of big data are required to be kept in the generation of synthetic data: (1) high volume and velocity data sets can be generated to support different benchmarking scenarios; (2) a diversity of data types and sources should be supported (variety); (3) the important characteristics of raw data should be preserved (veracity). To date, although some powerful data generators such as PDGF \cite{rabl2011data}, a parallel data generator designed to produce structured data, have been proposed, a majority of data generators in current big data benchmarks only focus on specific data types or targets one specific platform such as Hadoop. Moreover, little previous work pays attention to keeping  veracity of the real life data during data generation. In conclusion, current work has not adequately addressed issues relating to keeping the 4V properties of big data.

In this paper, we introduce Big Data Generator Suite (\emph{BDGS}), a comprehensive tool developed to generate synthetic big data while preserving 4V properties.
The data generators in BDGS are designed for a wide class of application domains (such as search engine, e-commence, and social network) in big data systems, and will be extended for other important domains. We demonstrate the effectiveness of BDGS by developing data generators based on six real data sets that cover three representative data types (structured, semi-structured,  and unstructured data), as well as three typical data sources (text, graph, and table). In any data generator, users can specify their preferred data volume and velocity in data generation. Theoretically, in BGDS, the volume and velocity of a data set can only be bounded by the storage capacity, the hardware configuration (e.g. the number of nodes), the parallelism strategy (how many generators execute in parallel), and the total execution time. At present, BDGS is implemented as a component of our open-source big data benchmarking project, \emph{BigDataBench} \cite{Bigdatabench,zhu2014gcbdw,luo2012cloudrank,jia2014implications,gao2013bigdatabench,jiacharacterization}, available at \url{http://prof.ict.ac.cn/BigDataBench}. BDGS supports all the 19 workloads and all six application scenarios in BigDataBench, including Micro Benchmarks, Basic Datastore Operations, Relational Query, Search Engine, Social Network and E-commerce. We also evaluate BDGS under different experimental settings. Our preliminary experimental results show that \emph{BDGS} can generate big data in a linear gross time as data volume increases while preserving the characteristics of real data.

The rest of the paper is organized as follows. Section \ref{Background} presents some basic concepts and requirements in big data generation. Section \ref{Related work} discusses related work. Section \ref{Methodology} provides an overview of our BDGS. Sections \ref{realdatasets} and \ref{synthetic_data_generation} explain how real-world data sets are selected and synthetic data sets are generated in \emph{BDGS}, respectively. Section \ref{Experimental Evaluation} reports our preliminary experimental evaluation of BDGS's effectiveness. Finally, Section \ref{conclusions} presents a summary and discusses directions for future work.
\vspace{-10pt}

\section{Background and Requirements} \label{Background}

Technically, a big data can be broken down into four dimensions: volume, variety, velocity, and veracity \cite{4v}, which form the 4V properties of big data.

\begin{enumerate}
  \item Volume is the most obvious feature of big data.
  With the exploding of data volume, we now use zettabytes (ZB) to measure data size, developed from petabytes (PB) measurement used just a short time ago.
  IDC forecasts a 44X increase in data volumes between 2009-2020, from 0.8 ZB to 35ZB. From recent Facebook's big data statistics, there are 500+ TB of new data generated per day \cite{carey2013bdms}.

  \item Velocity is another important but overlooked feature in big data, which denotes data generation, updating, or processing speed. The requirement of generating and processing data quickly gives rise to a great challenge to computer systems performance. According to Twitter's statistics, 340M tweets are sent out everyday, which means 3935 tweets are sent every second. 

  \item Variety refers to a variety of data types and sources. Comparing to traditional structured data in relational databases, most of today's data are unstructured or semi-structured and they are from different sources, such as movies' reviews, tweets, and photos.

  \item Veracity refers to the quality or credibility of the data. With the massive amount of data generated every day, e.g. there are 845M active users on Facebook and 140M active users on Twitter. How to derive trustworthy information from these users and discard noises, and how to generate realistic synthetic data on the basis of raw data, are still open questions.

\end{enumerate}

These properties of big data not only characterize features, but also bring requirements for a new generation of benchmarks. Based on the 4V properties of big data, we now define the requirements of data generation for benchmarking big data systems. Briefly, big data generators should scale up or down a synthetic data set (volume) of different types (variety) under a controllable generation rate (velocity) while keeping important characteristics of raw data (veracity).

\begin{enumerate}
  \item Volume: To benchmark a big data system, big data generators should be able to generate data whose volume ranges from GB to PB, and can also scale up and down data volumes to meet different testing requirements.

  \item Velocity: Big data applications can be divided into in three types: offline analytic, online service and realtime analytic. Generating input data of workloads to test different types of applications is the basic requirement for data generators. For applications of online services such as video streaming processing, data velocity also means the data processing speed. By contrast, for applications of offline analytic (e.g. k-means clustering or collaborative filtering) and realtime analytic (e.g. select or aggregate query in relational databases), data velocity denotes the data updating frequency. Big data generators should be able to control all the data velocities analyzed above.

  \item Variety: Since big data come from various workloads and systems, big data generators should support a diversity data types (structured, semi-structured and unstructured) and sources (table, text, graph, etc).

  \item Veracity: In synthetic data generation, the important characteristics of raw data must be preserved. The diversity of data types and sources as well as large data volumes bring a huge challenge in establishing trust in data generation.
      A possible solution is to apply state-of-the-art feature selection models, where each model is specialized for one type of data, to abstract important characteristics from raw data. The constructed models can then be used to generate synthetic data. We also note that in model construction or data generation, evaluation are needed to measure the conformity of the model or synthetic data to the raw data.
\end{enumerate}
\vspace{-18pt}


\section{Related work} \label{Related work}\vspace{-5pt}
How to obtain big data is an essential issue for big data benchmarking.
Margo Seltzer, et al \cite{seltzer1999case} pointed that if we want to produce performance results that are meaningful in the context of real applications, we need to use application-specific benchmarks. Application-specific benchmarking would require application-specific data generators which synthetically scaling up and down a synthetic data set and keeping this data set similar to real data \cite{tay2011data}. For big data benchmarking, we also need to generate data of different types and sources. We now review the data generation techniques in current big data benchmarks.


HiBench \cite{huang2010hibench} is a benchmark suite for Hadoop MapReduce. This benchmark contains four categories of workloads. The inputs of these workloads are either data sets of fixed size or scalable and synthetic data sets.


BigBench \cite{bigbench} is a recent effort towards designing big data benchmarks. The data generators in BigBench are developed based on PDGF \cite{rabl2011data}, which is a powerful data generator for structured data. In BigBench, PDGF is extended by adding a web log generator and a review generator.
But, in both generators, the veracity of logs and reviews rely on the table data generated by PDGF. In addition, the data generators in BigBench only support the workloads designed to test applications running in DBMSs and MapReduce systems.



Internet and industrial service providers also develop data generators to support their own benchmarks. In LinkBench \cite{armstrong2013linkbench}, the social graph data from Facebook are stored in MySQL databases. 
In this benchmark, the data generator is developed to generate synthetic data with similar characteristics to real social graph data. Specifically, the graph data are first broken into object and association types in order to describe a wide range of data types in the Facebook social graph.


The Transaction Processing Performance Council (TPC) proposes a series of benchmarks to test the performance of DBMSs in decision support systems. The TPC-DS \cite{tpcds} is TPC's latest decision support benchmark that implements a multi-dimensional data generator (MUDD). In MUDD, most of data are generated using traditional synthetic distributions such as a Gaussian distribution. However, for a small portion of crucial data sets, MUDD replies on real data sets to produce more realistic distributions in data generation. Although it can handle some aspects of big data such as volume and velocity, it is only designed for the structured data type.

\begin{table*}[!t]
\centering
\caption{Comparison of Big Data Benchmark's data generators}
\includegraphics[scale=0.6]{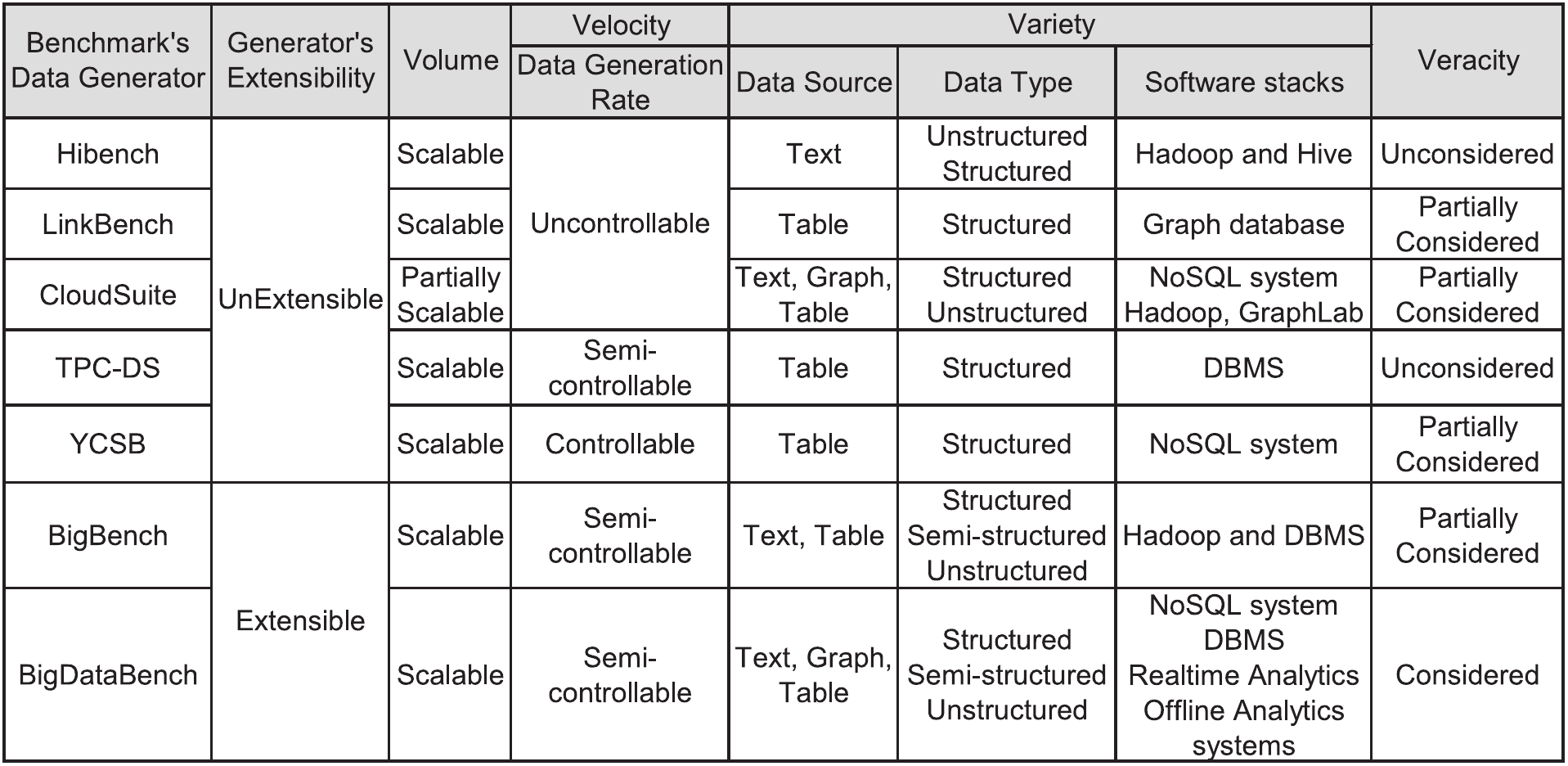}
\label{Relatedwork}\vspace{-20pt}
 \end{table*}

We summarize the data generation techniques in current benchmarks in Table \ref{Relatedwork}. Overall, existing benchmarks do not adequately address issues relating to keeping the 4V properties of big data. They are either designed for some specific application domains (e.g. the Hadoop system in Hibench, NoSQL system in YCSB, DBMSs in LinkBench and TPC-DS), or they only consider limited data types (e.g. the structured data type in YCSB and TPC-DS), which do not present the variety of big data. In addition,  fixed-size data sets are used as workload inputs in some benchmarks, we call the data volume in these benchmarks ``partially scalable".
Existing big data benchmarks rarely consider the control of data generation rates (velocity). Hence we call their data generators ``uncontrollable" or ``semi-controllable" in terms of data velocity. Our BDGS provides an enhanced control of data velocity by developing a mechanism to adjust data generation speeds.
Meanwhile, most of benchmarks employ random data sets as workload inputs, which do not consider the veracity of big data. In addition, some data generation tools use similar distributions with real scenarios (data veracity is partially considered).
Hence using such synthetic data, it is difficult to make accurate testing of realistic application scenarios; that is, current synthetic workload inputs incur a loss of data veracity.

\section{The Methodology of BDGS} \label{Methodology}

\begin{figure}[htb]
\centering
\includegraphics[scale=0.6]{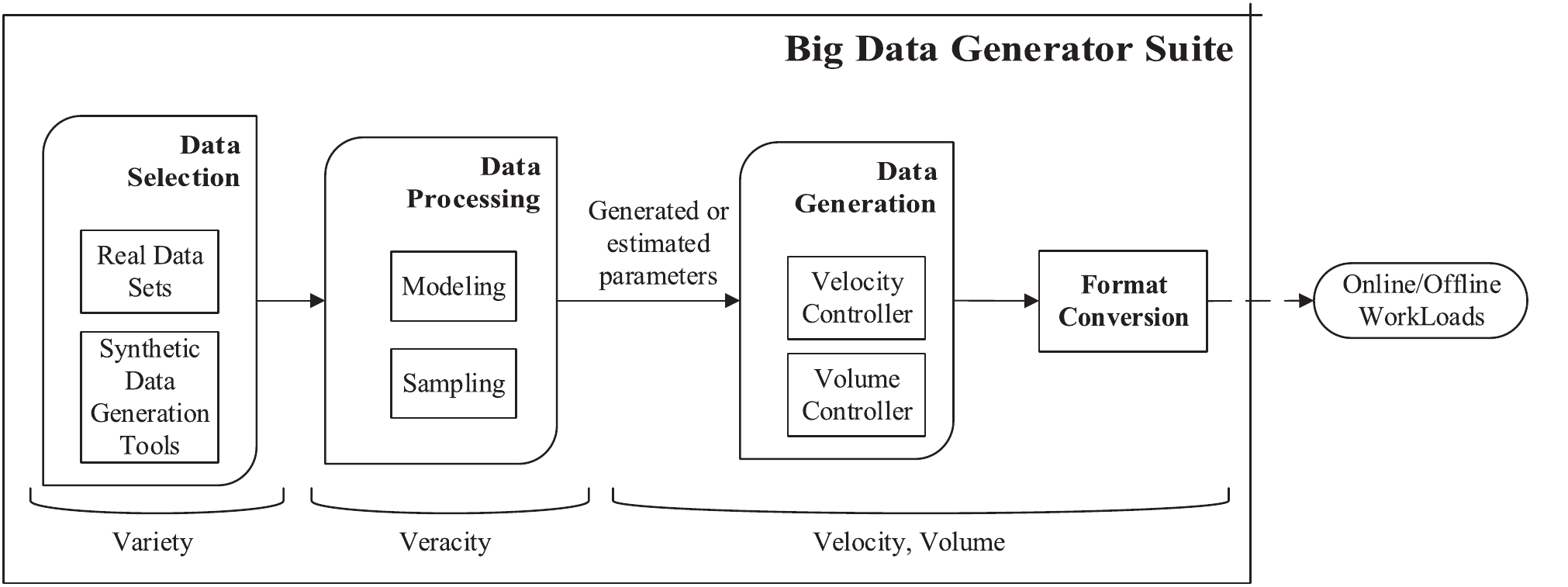}
\caption{The Architecture of BDGS}
\label{architecture}
\end{figure}

In this section, we present an overview of BDGS that is implemented as an component of our big data benchmark suite BigDataBench \cite{Bigdatabench}. Briefly, BDGS is designed to provide input data for application-specific workloads in BigDataBench. As shown in Figure \ref{architecture}, the data generation process in BDGS consists of four steps.

The first step is data selection, which reflects the variety of big data, in which the main operation is the selection of representative real-world data sets or tools to generate synthetic data sets.
The second step is data processing, the chosen data sets are processed to exact their important characteristics, thus preserving the veracity of data. Sampling methods are also provided to scale down data sets.
The third step is data generation, the volume and velocity are specified according to user requirements before a data set is generated.
Finally, the format conversion tools transform the generated data set into a format capable of being used as the input data of a specific workload. At present, we are working on implementing a parallel  version of BDGS.

\vspace{-15pt}
\vspace{-30pt}

\subsection{Selection of Data Sets}

In BDGS, we select the real data sets with the following considerations. First, real data sets should cover different data types including structured, semi-structured and unstructured data, as well as different data sources such as text data, graph data and table data. In addition, each chosen data set should be obtained from a typical application domain. Therefore, a generated synthetic data set is derived from a representative real data set, and can be used as the input of an application-specific workload. Section \ref{realdatasets} provides a detailed explanation of the selection of real data sets in BDGS.

\subsection{Generation of Synthetic Data Sets}

Overall, in BDGS, the method used to generate synthetic data sets is designed to keep the 4V properties of big data. First, the configuration of each data generator can be easily adjusted to produce different volumes of synthetic data sets. Similarly, different velocities of data sets can be generated by deploying different numbers of parallel data generators. In BDGS, different kinds of generators are developed to generate a variety of data sets with different types and sources. Finally, the synthetic data sets are generated using the data models derived from real data sets, thus preserving the data veracity in generation. Section \ref{synthetic_data_generation} provides a detailed explanation of the synthetic data generation in BDGS.

\vspace{-8pt}
\section{Real-world Data Sets} \label{realdatasets}\vspace{-8pt}

Our data generation tools are based on small-scale real data, this section introduces the representative real data sets we selected. To cover the variety and veracity, we choose six representative real data sets, which can be download from \url{http://prof.ict.ac.cn/BigDataBench/#downloads}.
As shown in Table \ref{realdata}, our chosen real data sets are diverse in three dimensions: data type, data source, and application domain. In the these data sets, the Wikipedia Entries, Google Web Graph, and Facebook Social Graph are unstructured; E-commence transaction data are structured; and Amazon Movie Reviews and Personal Resumes are semi-structured. Specifically, Wikipedia Entries are plaint text, and Amazon Movie Reviews are text but have some schema constrains. Google Web Graph are directed graphs in which nodes are web pages and each edge denotes a linking of two pages. Facebook Social Graph are undirected graphs in which two persons are friends of each other. As we shown later, Amazon Movie Reviews can be converted to bipartite graphs in which each edge has attributes like score and review text. E-commence transaction data are typical relational tables, and Personal Resumes can also be seen as table-like data with less schema. Finally, our chosen data sets are from different internet service domains: Wikipedia Entries and Google Web Graph can be used for the search engine domain, while Personal Resumes are from the vertical search engine. Amazon Movie Reviews and E-commence transaction are from the E-commence domain, and Facebook Social Graph are from the social network domain.

\vspace{-15pt}
\doublerulesep 0.1pt
\begin{table}[htb]
\center
\caption{The summary of six real life data sets. }\vspace{-15pt}
\label{realdata}
\begin{tabular}{|c|c|c|p{1.5in}|}

  \hline
  Data sets        &   Data type             & Data source    &   Data size \\
  \hline
  Wikipedia Entries   &   un-structured             & text data & 4,300,000 English articles\\
  \hline
  Amazon Movie Reviews   &   semi-structure           & text data         & 7,911,684 reviews \\
  \hline
  Google Web Graph      &   un-structured                 & graph data        &   875713 nodes, 5105039 edges  \\
  \hline
  Facebook Social Network    &   un-structured                 & graph data      & 4039 nodes, 88234 edges \\
  \hline
  E-commence Transaction       &   structured                   & table  data      & Table1: 4 columns, 38658 rows. Table2: 6 columns, 242735 rows  \\
  \hline
  Person Resumes Data     &   semi-structured            & table  data  &  278956 resumes\\
  \hline
\end{tabular}\vspace{-20pt}
\end{table}

\begin{enumerate}
\item \textbf{Wikipedia Entries} \cite{wikipedia}.
The Wikipedia data set is unstructured, with 4,300,000 English articles.

\item \textbf{Amazon Movie Reviews} \cite{amazonreview}. This data set is
semi-structured, consisting  of 7,911,684 reviews  on 889,176 movies
by 253,059 users. The data span from Aug 1997 to Oct 2012.
The raw format is text, and consists of productID, userID, profileName, helpfulness, score, time, summary and text.


\item \textbf{Google Web Graph} \cite{googleweb}. This data set is
unstructured, containing 875713  nodes representing web pages and
5105039 edges representing the links between web pages. This data
set is released  by Google as a part of Google Programming Contest.

\item \textbf{Facebook Social Graph} \cite{facebookgraph}. This data set
contains 4039 nodes, which represent users, and 88234 edges, which
represent friendship between users.

\item \textbf{E-commence Transaction}. This data set is from an E-commence
web site, which  is structured, consisting of two tables: ORDER
and order ITEM.  


\item \textbf{Personal Resumes}. This data is from a vertical search engine for scientists developed by ourselves.
The data set is semi-structured, consisting of 278956 resumes automatically extracted from 20,000,000 web pages of university and research institutions. The resume data have fields of name, email, telephone, address, date of birth, home place, institute, title, research interest, education experience, work experience, and publications. Because the data are automatically collected from the web by our program, they are not normalized: some information may be omitted in the web pages, while others may be redundant. For example, a person may only list name, email, institute, title while keeping other information blank, and he may have two emails.

\end{enumerate}

\vspace{-8pt}
\section{Synthetic Data Generators} \label{synthetic_data_generation}\vspace{-8pt}

This section presents our big data generator suite: BDGS, which is an implementation of our generic data generation approach in Figure 1. This implementation includes six data generators belonging to three types: Text Generator, Graph Generator and Table Generator.
BDGS can generate synthetic data while preserving the important characteristics of real data sets, can also rapidly scale data to meet the input requirements of all the 19 workloads in BigDataBench.
\vspace{-4pt}
\subsection{Text Generator}\vspace{-4pt}

Based on the chosen text data sets such as Wikipedia entries, we implement our text generator. It applies latent dirichlet allocation (LDA) \cite{blei2003latent} as the text data generation model. This model can keep the veracity of topic distributions in documents as well as the word distribution under each topic.

In machine learning and natural language processing, a topic model is a type of statistics designed to discover the abstract topics occurring in a collection of documents \cite{topicmodel}.
LDA models each document as a mixture of latent topics and a topic model is characterized by a distribution of words.
The document generation process in LDA has three steps:
\begin{enumerate}
  \item choose $N$ $\sim$ Poisson($\xi$) as the length of documents.
  \item choose $\theta$ $\sim$ Dirichlet($\alpha$) as  the topic proportions of document.
  \item for each of $N$ words $w_{n}$:

  (a) choose a topic $z_{n}$ $\sim$ Multinomial($\theta$)

    (b) choose a word $w_{n}$ from $p(w_{n}|,z_{n},\beta)$, a multinomial probability conditioned on the topic $z_{n}$
\end{enumerate}

\begin{figure}[htb]
\centering
\includegraphics[scale=0.65]{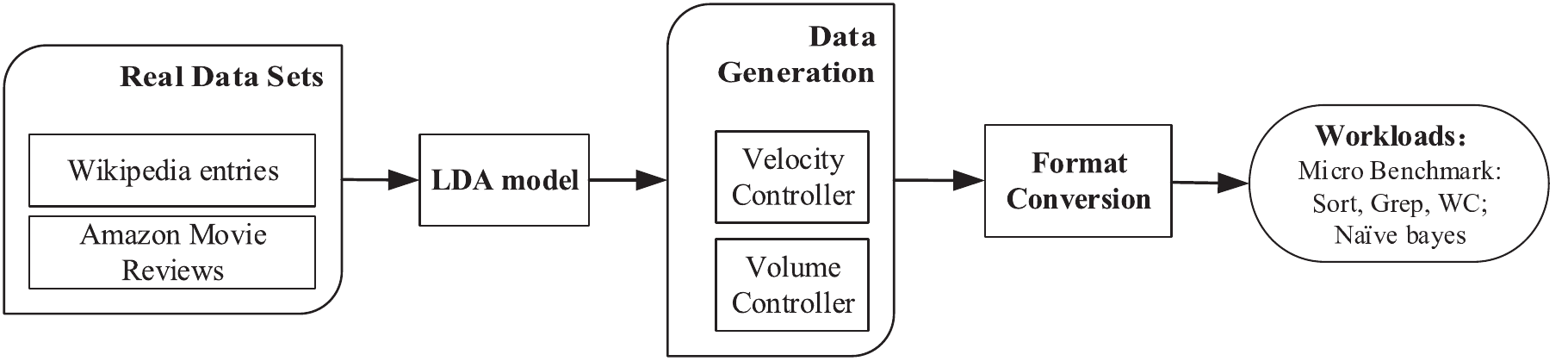}
\caption{The process of Text Generator}
\label{TextGen}\vspace{-10pt}
\end{figure}

Figure \ref{TextGen} shows the process of generating text data. It first preprocesses a real data set to obtain a word dictionary.
It then trains the parameters of a LDA model from this data set.
The parameters  $\alpha$ and $\beta$  are estimated using a variational EM algorithm, and the implementation of this algorithm is in lda-c \cite{lda-c}. Finally, we use the LDA process mentioned above  to generate documents.

Since LDA  models both word and hidden topics in text data, more characteristics of real raw data can be preserved than just applying the word-level modeling. Documents are independent of each other, so the generation of different documents can be paralleled and distributed using multiple generators under the same parameters, thus guaranteeing the velocity of generating data. Users can also configure the number of documents or the size of text data, thus generating different volumes of text data.

\subsection{Graph Generator}

A graph consists of nodes and edges that connect nodes. For the graph data sets, namely Facebook Social Graph and Google Web Gragh, we use the kronecker graph model in our graph generator. Using a small number of parameters, the kronecker graph model \cite{leskovec2005realistic}  can capture many graph patterns, e.g. the Denazification Power Law and the Shrinking Diameters effect.

Some temporal properties of real networks, such as densification and shrinking diameter can also be cover. We believe that the kronecker graph can effectively model the structure of real networks. Many temporal properties of real networks such as densification and shrinking diameter can also be covered by the kronecker graph \cite{leskovec2010kronecker}, thus guaranteeing the veracity of the generated graph data.

\begin{figure}[htb]
\centering
\includegraphics[scale=0.6]{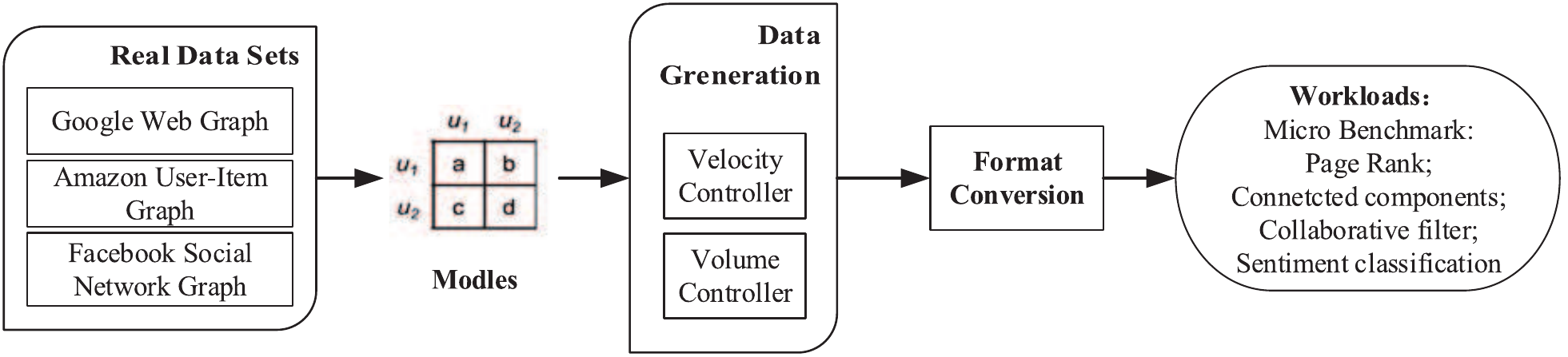}
\caption{The process of Graph Generator}
\label{GraphGen}\vspace{-20pt}
\end{figure}

The kronecker graph model is designed to create self-similar graphs. It begins with an initial graph, represented by adjacency matrix N.  It then progressively produces larger graphs by kronecher multiplication. Specifically, we use the algorithms in \cite{leskovec2010kronecker} to estimate initial $N$ as the parameter for the raw real graph data and use the library in Stanford Network Analysis Platform (SNAP, http://snap.stanford.edu/) to implement our generator.  Users can configure the number of nodes in the graphs, so the graph volume can be scaled up. The kronecker graph can be generated in linear time with the expected number of edges.  Figure \ref{GraphGen} shows the process of our Graph Generator.


%
%

\begin{figure}[htb]
\centering
\includegraphics[scale=0.6]{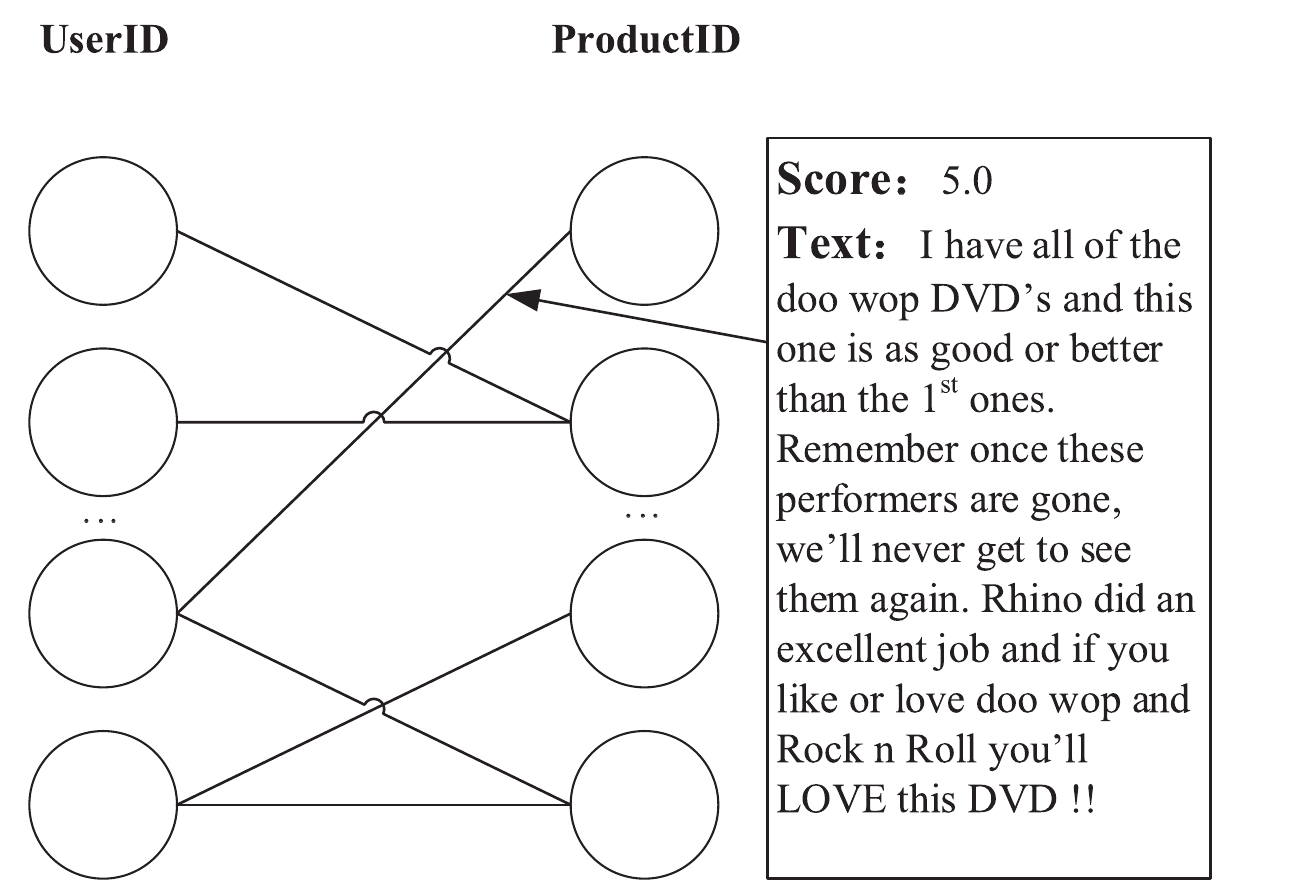}
\caption{The process of generating Review data}
\label{ReviewGen}\vspace{-20pt}
\end{figure}

Although the raw format is text, the Movie Reviews data set can also been seen as a graph.  As shown in Fig \ref{ReviewGen}, the left nodes represent users (each user has a Id), and the right nodes represent products (each product has a Id). The edge means the left user commented review on the right product, while the attributes of the edge are score and review text.

Movie Reviews data set can be used for two workloads: collaborative filtering and sentiment classification. For collaborative filtering, we use the productId, userId and score which can be seen as user-product score matrices. The task of collaborative filtering is to predict the missing score in the matrices.  While sentiment classification uses the review text as the input, and the score as the category label. As a result, when we generate the big synthetic review data,  we only concern the fields of productId, userId, score and text.


There are two steps to generate Movie Reviews data. First, applying graph generator to generate a bipartite graph. Second, for each edge in the graph, using the multinational random number generator to generate a score \emph{S}, then using the trained LDA under category S to generate review text for this edge.
\subsection{Table Generator}
To scale up the E-commence transection table data,  we use the PDGF \cite{rabl2011data}, which is also used in BigBench and TPC-DS.
PDGF uses XML configuration files for data description and distribution, thereby simplying the generation of different distributions of specified data sets. Hence we can easily use PDGF to generate synthetic table data by configuring the data schema and data distribution to adapt to real data.

Personal Resumes data are also a kind of table-like data (semi-structured data). We implement a resume generator to generate personal resumes for basic data store operations that are similar to those of YCSB. Traditionally, the input data in YCSB are table records that set all the fields as mandatory,
however, the real world data especially in NoSQL scenarios are schema less.  That is to say, the records can have arbitrary fields.
To model these uncertainties of real data, we use the professor resumes in ProfSearch (available at http://prof.ncic.ac.cn/)  as the real data set. A resume may consist of name, email, telephone, address, date of birth, home place, institute, title, research interest, education experience, work experience and publications.
Each resume's primary key is name, but other fields are optional. Moreover, some fields have sub fields. For example, experiences may have sub fields of time, company or school, position or degree, publication may have fields of author, time, title, and source.

\begin{figure}[htb]
\centering
\includegraphics[scale=0.6]{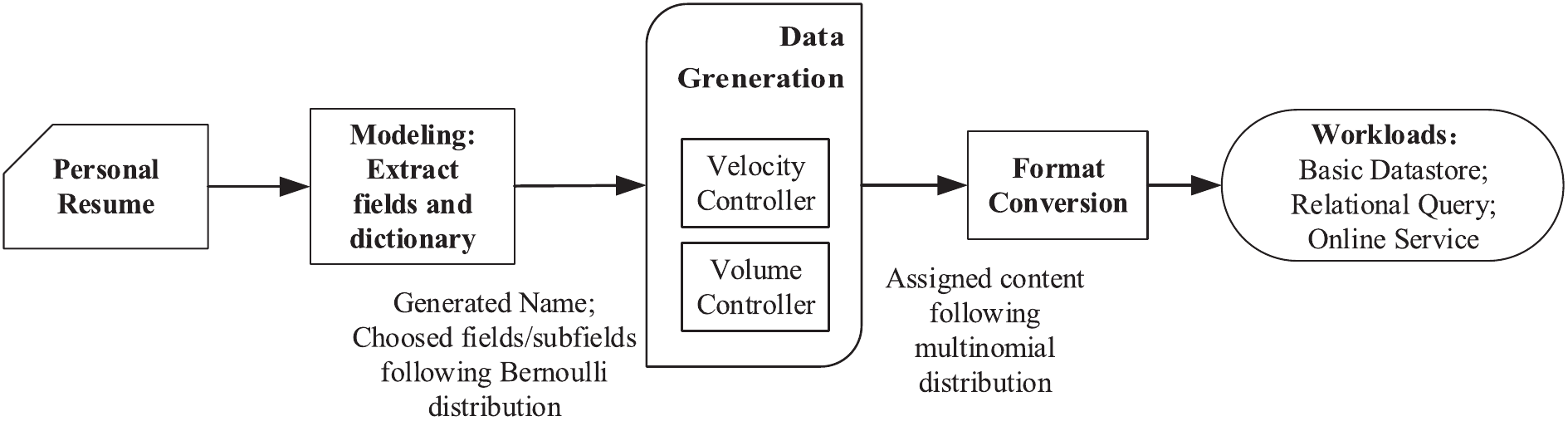}
\caption{The process of generating Resume data.}
\label{ProfGen}\vspace{-20pt}
\end{figure}

Since a field is a random variable whose value can be ¡°presence¡± or ¡°absence¡±, suppose the probability of the filed being ¡°presence¡± is p (0<=p<=1), then the probability of the opposite choice (the filed is ¡°absence¡±) is 1-p. Hence we can view the filed as a Binomial random variable and it meets the Binomial distribution. As a result, we generate personal resume data using the following three steps. Fig \ref{ProfGen} shows the process of resume generator.
\begin{enumerate}
\item Randomly generate a string as the name of a resume
\item Randomly choose fields from email, telephone, address, date of birth, home place, institute, title,
research interest, education experience, work experience and publications, where each field follows the bernoulli probability distribution.
\item For each field:

\textbf{if} the field has sub fields,  then randomly choose its sub fields following the bernoulli probability distribution.

\textbf{else} assign the content of the field using the multinomial probability distribution.
\end{enumerate}

\section{Experimental Evaluation} \label{Experimental Evaluation}

This section first describes the metric to evaluate our BDGS and the experimental configurations, following the results of experimental evaluation. The evaluation is designed to illustrate the effectiveness of our BDGS in generating different types of synthetic data sets based on real data sets from different sources. The evaluation results also shows the consumed generation time with respect to the volume of generated data.

We implemented BDGS in Linux environment, in which most of programs are written in C/C++ and shell scripts. Although the current experiments were doing on a single node, the generators and schemas can be deployed in multiple nodes.
\vspace{-8pt}
\subsection{Metric}
To evaluate the effectiveness of our methodology and our big data generator suite, we use a user-perceived performance metric--data generation rate to evaluate the data velocity under different data volumes. The data generation rate is defined as the amount of data generated divided by the running time for generation data. We use Edgs/s as the generation rate for graph data and MB/s as the generation rate for all the other data types. For examples, in graph data generation, generating 100000 edges in 100 seconds means the generation rate is 100000/100=1000 Edgs/s. In text data generation, generating 100 GB (i.e. 102400 MB) text data in 10000 seconds means the generation rate is 102400/10000=10.24MB/s.

\subsection{Experiment Configurations}
We ran a series of data generating experiments using BDGS, in which, currently, we choose our Text, Graph and Table Generators to generate text, graph, and table data, respectively. We use two real data sets: Wikipedia Entries and Amazon Movie Reviews for the Text Generator; Amazon Movie Reviews's graph data, Facebook data set and Google data set for the Graph Generator; and E-commerce Transaction Data for the Table Generator.

In experiments, we generated big data on one node with two Xeon E5645 processors equipped with 32GB memory and 8X1TB disk.
In generation, the data size ranges from 10 GB to 500 GB for the Text Generator and the Table Generator, and $2^{16}$ to $2^{20}$ for the Graph Generator.


\subsection{Evaluation of BDGS}

\begin{figure}[htb]
\centering
\includegraphics[scale=0.6]{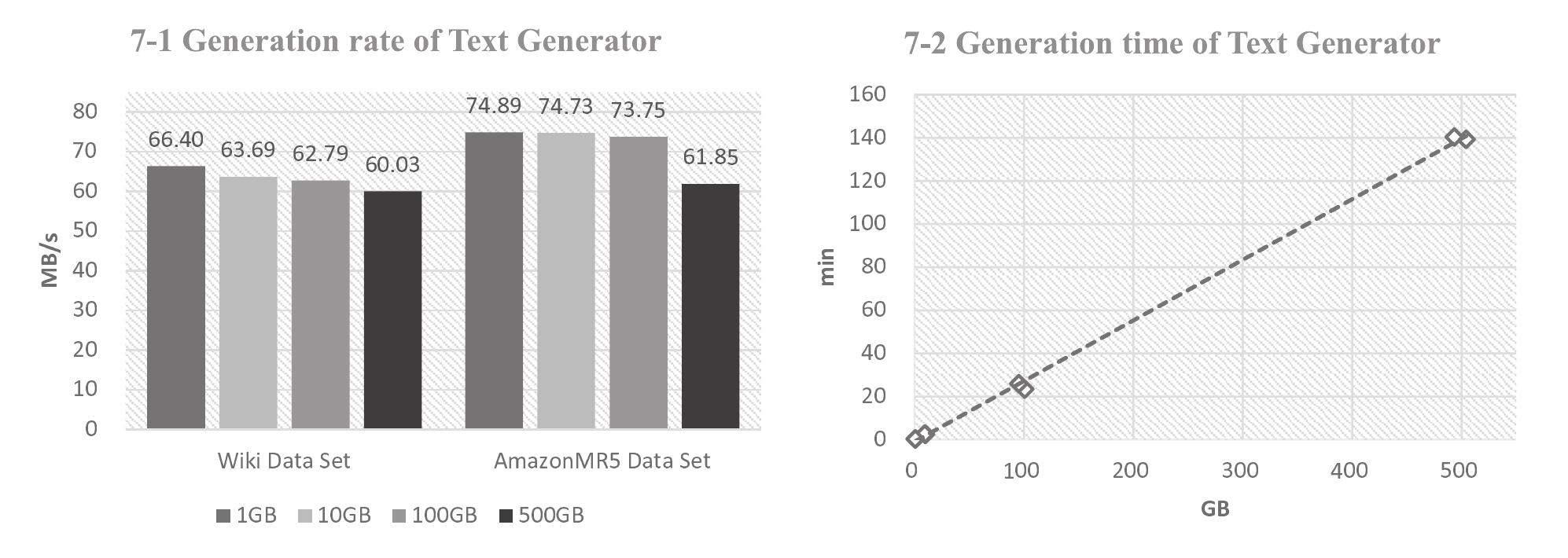}
\caption{Data Generation Rates and Time of the Text Generator}
\label{exTextGen}
\end{figure}

\begin{figure}[htb]
\centering
\includegraphics[scale=0.6]{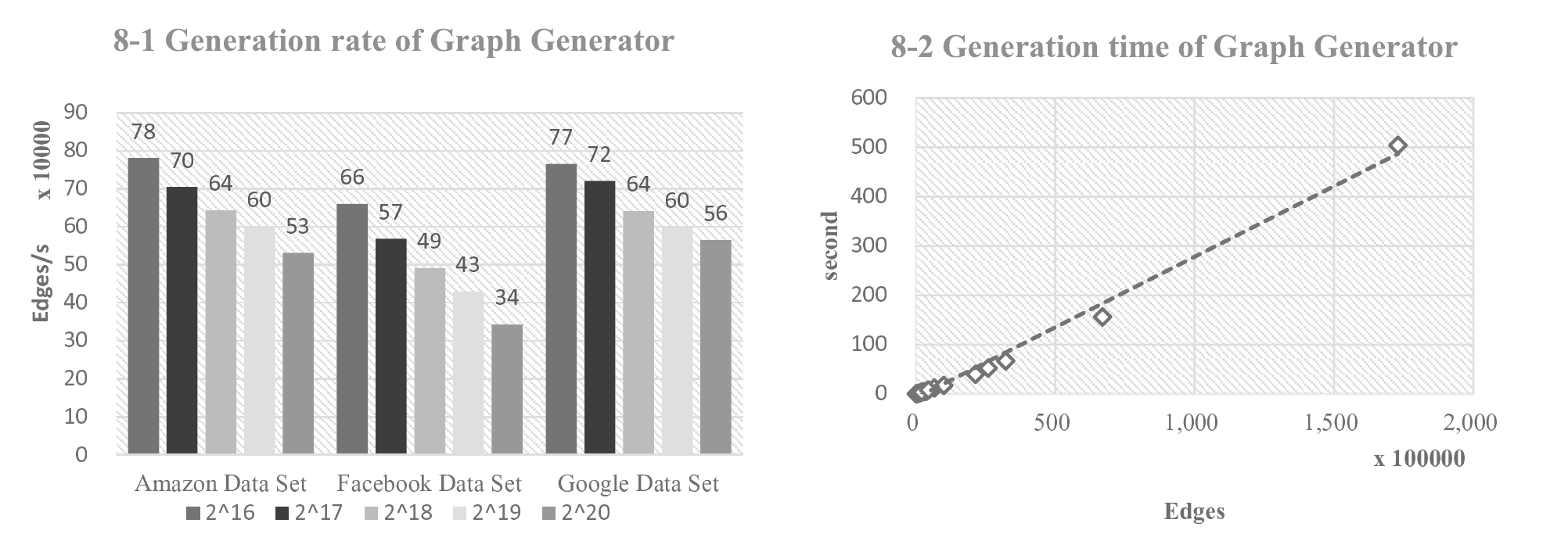}
\caption{Data Generation Rates and Time of the Graph Generator}
\label{exGraphGen}
\end{figure}

\begin{figure}[htb]
\centering
\includegraphics[scale=0.6]{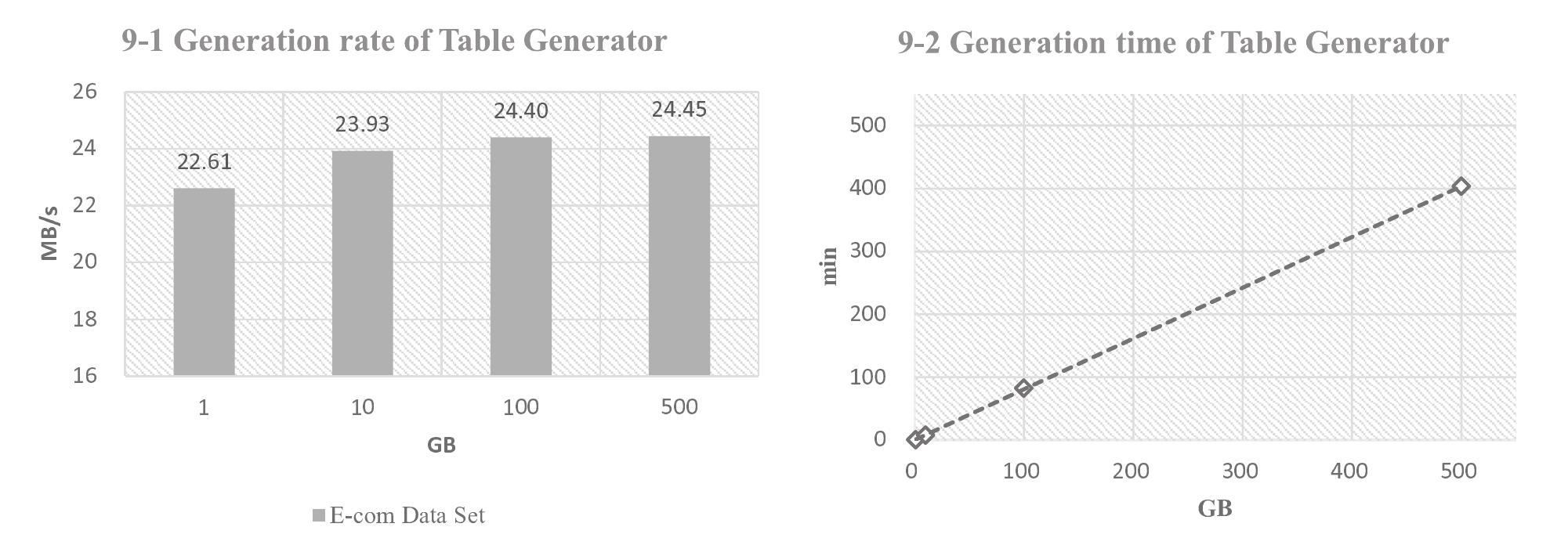}
\caption{Data Generation Rates and Time of the Table Generator}
\label{exTableGen}
\end{figure}

We tested our Text, Graph and Table Generators under different data volumes and employed the data generation rate to evaluate these generators. Figures \ref{exTextGen}-1, \ref{exGraphGen}-1 and \ref{exTableGen}-1 report the data generation rates of the Text, Graph and Table Generator, respectively. In the Text Generator, the average generation rate of the Wiki data set is 63.23 MB/s, and this rate of Amazon movies reviews(5 scores) data set is 71.3 MB/s.
The difference is caused by the dictionary's size, which is 7762 in Wiki data set, and 5390 in Amazon data set. The least average generation rate of the Graph Generator under enough memory is 591684 Edges/s, and this rate of the Table Generator is 23.85 MB/s.

Furthermore, the experiment results in Figures \ref{exTextGen}-2, \ref{exGraphGen}-2 and \ref{exTableGen}-2 show that our generators can rapidly generate big data in a linear gross time with the data volume increases. Hence, it is reasonable to say that our generators can maintain a roughly constant rate to generate high volume data for different data types.
For example, generating 1 TB Wiki data takes 4.7 hours.

In addition, we can observe in Figures 7, 8 and 9 that although the data generation rates roughly maintain constants, these rates have slightly changes when dealing with different data types. For the Text Generator and the Graph Generator, the data generation rate varies slightly in spite of which real data sets are used, and this variation is mainly caused by the high volume of data and the limitation of memory capacity.
Based on the observation that the memory resource utilization is above 90\% in experiments, we believe the performance bottleneck of the Text Generator is memory.
The next edition should pay more attention to optimize memory control. In addition, since the Graph Generator needs to compute the whole Graph in memory to match the real data sets and generate a big map, the Graph Generator has a larger memory requirement. This incurs the smallest data generation rate in the Graph Generator.
Finally, in the Table Generator (Figure 9), the data generation rate slightly increases with the data volume; that is, the average execution time to generate a unit of data decreases. This is because the total execution time of the Table Generator consists of a long configuration time and the data generation time. Given that the generation time per unit of data keeps fixed when the data volume increases, the average configuration time per unit of data  (i.e. the total configuration time divided by the total data size) deceases as data volume increases. Therefore, the execution time per unit of data (i.e. the sum of the data generation time per unit of data and the average configuration time per unit of data) decreases.


To summarize, our big data generator suite can rapidly generate big data in a linear gross time as data volume increases. It not only covers various data types and sources, but also considers generating big data on the basis of representative real data sets, thereby preserving the characteristics of real data.

\section{Conclusions} \label{conclusions}\vspace{-8pt}
In this paper, we argued that evaluation of big data systems and the diversity of data types raises new challenges for generating big data with 4V properties in benchmarks. Based on this argument, we proposed a tool, BDGS, that provides a generic approach to preserve the 4V properties in the generation of synthetic data. Our BDGS covers three representative data types including structured, semi-structured and unstructured, as well as a wide range of application domains including search engine, social network, and electronic commerce. To demonstrate the effectiveness of BDGS, three kinds of generators together with their data format conversion tools were developed and experimentally evaluated. The experiment results show that our BDGS can rapidly generate big data in a linear gross time as data volume increases.

The approach presented in this paper provides a foundation to develop data generation tools for a wide range of big data applications.
At present, we are working on implementing a parallel version of BDGS and testing its performance on veracity characteristic.
In the future, we plan to investigate the quality of synthetic data, which is decided by the statistical characteristic of both real and synthetic data as well as the workload performance. We will also add more data sources such as multimedia data to our BDGS.
We will also work on applying the same data generator in different workloads, in which the implementation of big data applications and systems varies.

%
%
%
%
%
%
%
%

\end{spacing}
\end{document}